\documentclass[printer]{aa}
\usepackage[latin1]{inputenc}
\usepackage[T1]{fontenc}
\usepackage[english]{babel}
\usepackage{txfonts}
\usepackage{latexsym}
\usepackage{graphicx}
\usepackage[square]{natbib} 
\usepackage{color}
\bibpunct{(}{)}{;}{a}{}{,}
\begin{document}
\title{Large-scale structure and the Cardassian fluid}
\author{Stéphane Fay\inst{1}\inst{2}\thanks{steph.fay@gmail.com} \and Morad Amarzguioui\inst{3}\thanks{morad@astro.uio.no}}
\institute{
School of Mathematical Science,\\
Queen Mary, University of London, Mile End road, London E1 4NS, UK
\and 
Laboratoire Univers et Théories (LUTH), UMR 8102\\
Observatoire de Paris, F-92195 Meudon Cedex, France
\and
Institute of Theoretical Astrophysics,\\
University of Oslo, PO Box 1029 Blindern, 0315 Oslo, Norway
}
\authorrunning{S. Fay \& M. Amarzguioui}
\titlerunning{Large-scale structure and the Cardassian fluid}
\abstract{
In this paper, we confront the predictions of the
power law cardassian model for the baryon power
spectrum with the observations of the SDSS galaxy survey. We show that
they fit only for very unusual values of the cold dark matter or
baryon density parameters, the Hubble parameter or the spectral index
of the initial power spectrum. 
Moreover, the best-fit Cardassian models turn out to
  be phanton models. If one wants to recover the usual values for these constants, as
quoted by the WMAP team, the power law Cardassian model turns out to
be indistinguishable from a $\Lambda$CDM model. 
\keywords{Theory -- Cosmological parameters -- Large-scale structure of Universe}
}
\maketitle
\section{Introduction} \label{s0}
One of the greatest discoveries in cosmology in recent years is the accelerated expansion of the Universe. The first strong evidence that led to the conclusion that the expansion of the Universe appears to be accelerating, came in 1998 from observations of supernovae of type Ia \citep{Riess98, Per99}. Since then, more recent supernovae observations \citep{Knop03, Tonry03, Rie04} along with observations of the cosmic microwave background \citep{Spergel03} and large-scale structure \citep{Tegmark03a} have strengthened this conclusion even further. 

Now that the accelerated expansion of the Universe seems to have been
established, the task facing cosmologists is to explain its origin.
Several models have been proposed over the years that attempt to
explain the \emph{dark energy} that gives rise to the accelerated
expansion. One such model is the power law Cardassian
\citep{Freese02}. This is a model that has no additional energy
components other than baryons and cold dark matter. However, motivated
by brane world cosmology \citep{Chung99}, the Friedmann equation is
modified by the presence of an additional energy term that is
proportional to the cold dark matter density raised to a general
power, i.e. $\rho_m^n$. It is this term that gives rise to the
accelerated expansion or dark energy. Indeed, it can
  be shown that the Hubble function can be written in the same form as
  ordinary general relativity with a dark energy fluid
  defined by a constant equation of state $p_{DE}/\rho_{DE}=n-1$,
  where $p_{DE}$ and $\rho_{DE}$ are respectively the pressure and
  density of this special form of dark energy \citep{Avelino02}.
  Consequently, acceleration occurs for $n<2/3$ and the power law
  Cardassian model cannot be distinguished from such dark energy
  models by any observational tests based on the Hubble function only,
  such as the redshift-luminosity distance relation inferred from
  supernovae. However, this is no longer the case for tests based on
  large-scale structure data \citep{Sandvik02,Koivisto05}, since they
  probe more than just the background evolution.
More general Cardassian models exist where the
  modification of the Friedmann equation cannot be written as a simple
  power in the matter density. Examples of such model are the
  Polytropic Cardassian Model (PC) and the Modified Polytropic
  Cardassian Model (MPC), which were proposed in \citet{Gondolo02}.
  However, for such models it is more difficult to make a connection
  to fundamental physics, hence rendering the model a purely
  phenomenological model with little physical motivation. We stress
  that we will consider only the original power law Cardassian model
  in this work. Furthermore, the model will be considered in the
  so-called fluid interpretation \citep{Gondolo02}, which will be
  explained in section~\ref{s1}. In the following, we will refer to
  the the power law Cardassian model as a \emph{phantom Cardassian}
  when $n<0$ 

Constraints on Cardassian model imposed by the supernovae have been
investigated extensively in several papers, e.g. \citep{Sen02, Wang03,
  Frith03, Gong04, Nesseris04, Zhu04, Bento05, Bento06,
  Lazkoz05, Szyd04, Szyd05}. These studies show that
the parameter space of the model is quite degenerate. In order to get
better constraints one needs to consider additional cosmological
tests. In \cite{AmaElgMul04}, the MPC model is constrained by
comparing the matter power spectrum predicted by the model with that
inferred from the SDSS large-scale structure data. The authors show
that these data constrain the model severely. In fact, it is shown
that $|n|$ has to be of order less than $10^{-5}$, which renders the
model virtually indistinguishable from the $\Lambda$CDM model. In the
analysis leading to this conclusion the density and Hubble parameters
were fixed to the first-year values quoted by the WMAP team
\citep{Spergel03}, i.e. $\Omega_{m0}=0.224$, $\Omega_{b0}=0.046$ and
$h=0.72$. Here, $\Omega_{m0}$ and $\Omega_{b0}$ denote the density
parameters of dark matter and baryons respectively. In this paper we
want to generalise this approach. We restrict
  ourselves to the power law Cardassian model, and perform a fit to
  the baryon power spectrum while allowing these cosmological
  parameters to take a range of constant values along with $n$.
Furthermore, we allow also the spectral index $n_s$ of the initial
power spectrum to take a range of possible values
  rather than being just unity. This makes sense in view of the last
  WMAP results \citep{Spergel06} which give $n_s=0.95\pm 0.016$
  compared with the first-year estimate of $0.99\pm0.04$. Does this
added freedom allow the Cardassian model to deviate from the
$\Lambda$CDM model? This is what we intend to answer in this paper.

We find that all the best-fit models have negative
  $n$, which in the terminology defined above means that they are
  phantom models. The models that are compatible with the data fall
  into two categories: $|n|$ is either so small that the model is
  indistinguishable from a $\Lambda$CDM model, or $n$ can have a
  non-negligible negative value. But in the latter case, one finds
  that either the cold dark matter density, the Hubble parameter or
  the spectral index must have very unusual values for the model to
  differ substantially from a $\Lambda$CDM model. Thus, if one wants
  to recover the usual values for these constants, as quoted by the
  WMAP team, the Cardassian model turns out to be indistinguishable
  from a $\Lambda$CDM model with $\mid n \mid<10^{-5}$.

The structure of this paper is as follows: In section~\ref{s1} we
present the Friedmann equation of the power law
Cardassian model and look at how first order perturbations evolve in
the fluid interpretation of this model. In section~\ref{s2} we discuss
how to obtain the baryon power spectrum for the Cardassian model using
the perturbed equations presented in the preceding section, and what
initial condition to use for these. In section~\ref{s3} we present the
statistics we need to fit the predicted power spectrum of the model
with the observed. In section~\ref{s4}, we fit the predictions of the
Cardassian model with the SDSS \citep{Teg03} baryonic power spectrum
by keeping three of the parameters $\Omega_{m0},\Omega_{b0},h,n_s$ fixed
to their WMAP concordance values and allowing the remaining one
to take a range of possible values in addition to $n$. 

Finally, in section~\ref{s5} we summarise and conclude.
\section{Field equations} \label{s1}
We assume the universe to be homogeneous, isotropic and flat. The
metric can then be written in the usual form: 
\begin{equation} \label{FRWmetric}
ds^2=-dt^2+a(t)^2(dr^2+r^2(d\theta^2+\sin^2\!\theta d\phi^2))
\end{equation}
The modified Friedmann equation for the Cardassian model is
\begin{equation}\label{fri1}
H^2=\frac{8\pi G}{3}(\rho_m +\rho_b+\rho_c)\,,
\end{equation}
where $\rho_m$ and $\rho_b$ are respectively the CDM and baryonic matter densities and $\rho_c\sim\rho_m^n$. Note that we are looking at the ordinary Cardassian model and not the Modified Polytropic Cardassian Model (MPC) that was considered in \cite{AmaElgMul04}. 

Energy conservation of pressureless matter implies that $\rho_m,\rho_b\sim(1+z)^3$ and hence, $\rho_c\sim(1+z)^{3n}$. This allows us to write the Friedmann equation as
\begin{equation}\label{fri2}
(\frac{H}{H_0})^2=(\Omega_{m0} +\Omega_{b0}) (1+z)^3+\Omega_{c0} (1+z)^{3n}\,,
\end{equation}
where we have defined the density parameters
\begin{equation} \label{densparams}
\Omega_{m0}=\frac{8\pi G\rho_{m0}}{3H_0^2},\quad
\Omega_{b0}=\frac{8\pi G\rho_{b0}}{3H_0^2},\quad
\Omega_{c0}=\frac{8\pi G\rho_{c0}}{3H_0^2}\,.
\end{equation}
In order to derive the matter power spectrum we need to go one step further than Eq.~(\ref{fri2}) which gives the evolution of the background. We need to consider first order perturbations. In order to do this, we follow the approach of \cite{AmaElgMul04} and consider the fluid interpretation of the Cardassian model as introduced in \cite{Gondolo02}. In this approach, the dark matter term and the term that gives rise to dark energy are treated as two "components" of one single fluid -- the Cardassian fluid. We write
\begin{equation} \label{defofrho}
\rho = \rho_m + \rho_c = \rho_m + \frac{\Omega_{c0}}{\Omega_{m0}}\rho_{m0}^{1-n}\rho_m^n \,.
\end{equation}
Treating this as a perfect fluid and using energy-momentum conservation, we arrive at the usual continuity equation for the fluid $\rho$:
\begin{equation} \label{conteq}
\dot{\rho}+3H(\rho+p)=0\,.
\end{equation}
Substituting from Eq.~(\ref{defofrho}) and using energy conservation of ordinary matter, the effective pressure of the Cardassian fluid can be written as
\begin{equation} \label{effp}
p = p(\rho_m) = \rho_m\frac{d\rho}{d\rho_m}-\rho\,.
\end{equation}
The sound speed and the effective equation of state for the Cardassian fluid can now be calculated easily. We get
\begin{eqnarray} \label{cs2}
c_s^2&=&\frac{\partial p}{\partial \rho}=\rho_m\frac{d^2\rho/d\rho_m^2}{d\rho/d\rho_m}\nonumber\\
&=&\frac{n(n-1)(\Omega_{m0}+\Omega_{b0}-1)(1+z)^{-3}}{n(\Omega_{m0}+\Omega_{b0}-1)(1+z)^{-3}-\Omega_{m0}(1+z)^{-3n}}
\end{eqnarray}
for the sound speed, and
\begin{eqnarray} \label{EOS}
\omega&=&\rho_m\frac{d\rho/d\rho_m}{\rho}-1\nonumber\\
&=&\frac{(n-1)(\Omega_{m0}+\Omega_{b0}-1)(1+z)^{-3}}{(\Omega_{m0}+\Omega_{b0}-1)(1+z)^{-3}-\Omega_{m0}(1+z)^{-3n}}
\end{eqnarray}
for the equation of state.

In the fluid interpretation, the model we're considering contains two fluids, namely the Cardassian fluid and the baryon fluid. The evolution of density perturbation in these fluids is given by a coupled set of second order differential equations. The detailed derivation of these equations can be found in \cite{LytSte90} and \cite{Pad93}. Setting the equation of state parameter and sound speed of baryons equal to zero, we can write these differential equations as
\begin{eqnarray} \label{DEcard0}
&&\ddot{\delta}+H\left(2-3(2w-c_s^2)\right)\dot{\delta}-\left(\frac{3}{2}H^2(7w-3w^2-6c_s^2)-(\frac{c_sk}{a})^2\right)\delta\nonumber\\
&&-\frac{3}{2}H^2(1+w)\frac{\rho\delta+\rho_b\delta_b}{\rho+\rho_b}=0
\end{eqnarray}
and
\begin{equation} \label{DEbar0}
\ddot{\delta}_b+2H\dot{\delta}_b-\frac{3}{2}H^2\frac{\rho\delta+\rho_b\delta_b}{\rho+\rho_b}=0\,,
\end{equation}
where $\delta$ and $\delta_b$ are the perturbations in the Cardassian and baryon fluids respectively. These equations are written with respect to cosmic time $t$. We want to solve them numerically, and in that respect it is more useful to write them in terms of a new time variable $u$ defined by
\begin{equation} \label{defofu}
\frac{d}{dt}=H\frac{d}{du}\,.
\end{equation}
This implies that $a=e^u=1/(1+z)$ if we define the scale factor today as $a_0=1$. Using this time parameter and marking a derivative with respect to it with a prime, the differential equations take the form
\begin{eqnarray} \label{DEcard}
&&H^2\delta''+(H^2)' \frac{\delta'}{2}+\left[2-3(2\omega-c_s^2)\right]H^2\delta'-\frac{3}{2}(7\omega-3\omega^2-\nonumber\\
&&6c_s^2)H^2\delta+\frac{k^2}{e^{u}}c_s^2\delta-\frac{3}{2}H^2(1+\omega)\frac{\rho\delta+\rho_b\delta_b}{\rho+\rho_b}=0
\end{eqnarray}
\begin{equation} \label{DEbar}
H^2\delta_b'' +(H^2)'\frac{\delta_b'}{2}-\frac{3}{2}H^2 \frac{\rho\delta+\rho_b\delta_b}{\rho+\rho_b}=0
\end{equation}
\section{Power spectrum} \label{s2}
\begin{figure}[h]
\resizebox{\hsize}{!}{\includegraphics{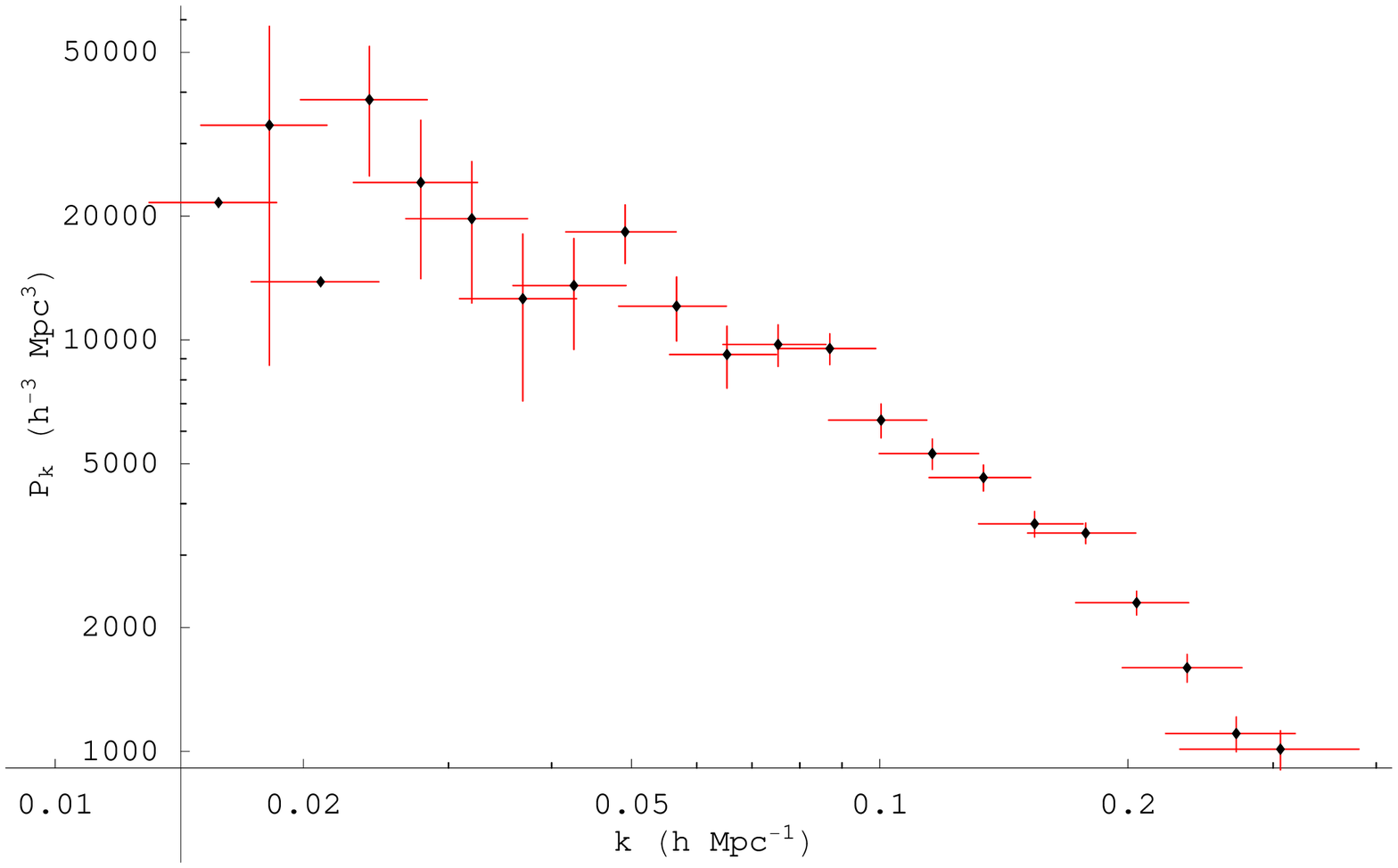}}
\caption{\label{dataSDSS}The 22 data points of the SDSS power spectrum}
\end{figure}

In this work we will use the large-scale structure data of the SDSS team to constrain the Cardassian model. The galaxy power spectrum inferred from these data\footnote{http://www.hep.upenn.edu/$\sim$max/sdss.html} is plotted in Fig.~\ref{dataSDSS}.

The power spectrum of energy component $i$ can be written as $P_i(k)\sim\delta_i(k)^2$. Thus, we can use Eqs. (\ref{DEcard}) and (\ref{DEbar}) with the appropriate initial conditions to predict the power spectrum of the baryons and/or the Cardassian fluid today. Taking the arguments presented in \cite{Beca03} into account, we will consider the power spectrum arising from the baryons only and not from the dark matter component of the Cardassian fluid.

To determine the initial conditions for the coupled system of
differential equations (\ref{DEcard}) and (\ref{DEbar}), we follow the
approach of \cite{Sandvik02} and \cite{AmaElgMul04}: we assume the
evolution of the Cardassian and $\Lambda$CDM models to be similar very
early on, and use CMBFast \citep{SelZal96} to calculate the matter transfer function at redshift $z=200$ arising from an initial spectrum with a general spectral index $n_s$. This allows us to write the density perturbations at $z=200$ as
\begin{equation} \label{initdens}
\delta(k,z=200)\sim k^{n_s/2}T(z,k)\,
\end{equation}
where $T(z,k)$ is the transfer function obtained from CMBFast. Next, we use (\ref{DEcard}) and (\ref{DEbar}) to evolve the perturbations until today. The power spectrum of baryons today is then obtained as $P_k\sim\delta_b(k,z=0)^2$.

\section{Statistics} \label{s3}
To fit the $22$ SDSS data points for the power spectrum we will use a least square test $\chi^2$. It is defined by
\begin{equation} \label{defofchi2}
\chi^2=\sum_{i=1}^{22}\frac{(P(k,0)^{\mbox{obs}}_i-P(k,0)^{\mbox{th}}_i N)^2}{\sigma^2_i}\,,
\end{equation}
where $P(k,0)$ is the power spectrum of the baryons today and the superscripts \emph{th} and \emph{obs} denote the \emph{theoretical} and the \emph{observed} spectra respectively. Furthermore, $\sigma_i$ are the $1\,\sigma$ errors of the SDSS data and $N$ is a normalisation factor that allows us to adjust the amplitude of the theoretical power spectrum. We want to find the values of $N$ and the other cosmological parameters that minimise $\chi^2$. To simplify this procedure, we can define a new quantity $\bar\chi^2$ which minimises $\chi^2$ analytically with respect to the amplitude $N$. First, we define the following quantities:
\begin{eqnarray}
&&A=\sum_{i=1}^{22}\frac{(P(k,0)_i^{\mbox{th}})^2}{\sigma^2_i}\\
&&B=\sum_{i=1}^{22}\frac{P(k,0)_i^{\mbox{obs}}P(k,0)^{\mbox{th}}_i}{\sigma^2_i}\\
&&C=\sum_{i=1}^{22}\frac{(P(k,0)_i^{\mbox{obs}})^2}{\sigma^2_i}
\end{eqnarray}
Substituting these into (\ref{defofchi2}), we get the expression
\begin{equation} \label{chi2_2}
\chi^2=AN^2-2BN+C
\end{equation}
Treating this as a function of $N$ we find that its minimum is reached for $N=B/A$. We define the value of $\chi^2$ corresponding to this minimising $N$ as the new $\bar\chi^2$. Thus, we write
\begin{equation} \label{barchi2}
\bar\chi^2=-\frac{B^2}{A}+C
\end{equation}
This last expression defines a new $\chi^2$ analytically minimised with respect to $N$. This is the definition of $\chi^2$ that we will use to fit the data.
\section{Fit to the SDSS matter power spectrum} \label{s4}
In this section we perform the fitting of the
  theoretical Cardassian model to the empirical values of the SDSS
  matter power spectrum. The parameters which are constrained in the
  fitting are the Cardassian parameter $n$, the CDM density parameter
  $\Omega_{m0}$, the baryon density parameter $\Omega_{b0}$, the
  Hubble parameter $h$, and the spectral index $n_s$. Ideally we would
  like to get a constraint on all of these parameters
  simultaneously, but the parameter space would be too large to handle
  in a simple $\chi^2$ test. We will therefore restrict ourselves to
  constraining pairs of these parameters with one parameter being
  $n$.
\subsection{Constraining the $n$ parameter only} \label{s41}
\begin{figure}[h]
\resizebox{\hsize}{!}{\includegraphics{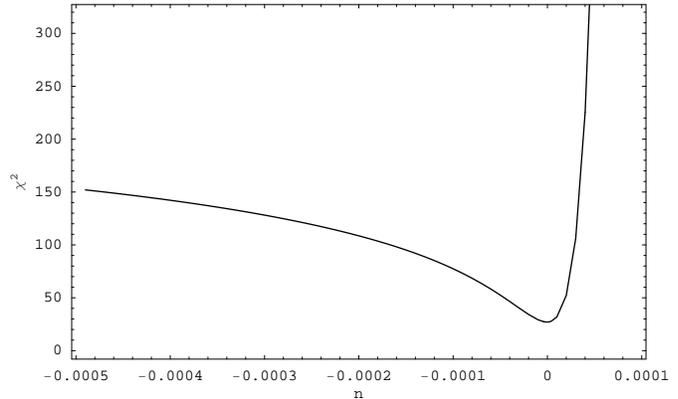}}
\caption{\label{card(n)chi2}$\chi^2$ as a function of $n$}
\end{figure}

In this first subsection we mainly try to recover the results of
\cite{AmaElgMul04}. In their analysis, the cosmological parameters
were set equal to the values quoted by the WMAP team. However, they
did not perform an explicit $\chi^2$ fitting to the data. They
looked at the predicted power spectrum and noted that it deviates
strongly from the SDSS data even for extremely small deviation from
the $\Lambda$CDM model. Repeating their analysis, but doing an
explicit fit to the data, we set $\Omega_{m0}=0.224$ and
$\Omega_{b0}=0.046$, implying $\Omega_{c0}=0.73$. Furthermore, we set
the value of the Hubble parameter to $h=0.72$ and the spectral index
to $n_s=1$. Using Eq. (\ref{barchi2}), we find the minimal value for
$\chi^2$ to be $27.08$ when $n=-5.6\times 10^{-7}$. With $22$ data
points and one free parameter, the $\chi^2$ per degree of freedom
(DOF) is thus $\chi^2_{DOF}=1.28$. For $n=0$, corresponding to a
cosmological constant, $\chi^2_{DOF}=1.29$. These values are not very
good but they are comparable to the values of $\chi^2_{DOF}$ that were
obtained with the $56$ supernovae of Perlmutter \citep{Per99}
published in $1999$, and in works aiming to constrain quintessence
models \citep{PieCla03}. So maybe more precise data in the future will
decrease the $\chi^2_{DOF}$ we get in this section, just as was the
case with the supernovae data. Figure~\ref{card(n)chi2} shows a plot
of $\chi^2$ as a function of the parameter $n$.  

\begin{figure}[h]
\resizebox{\hsize}{!}{\includegraphics{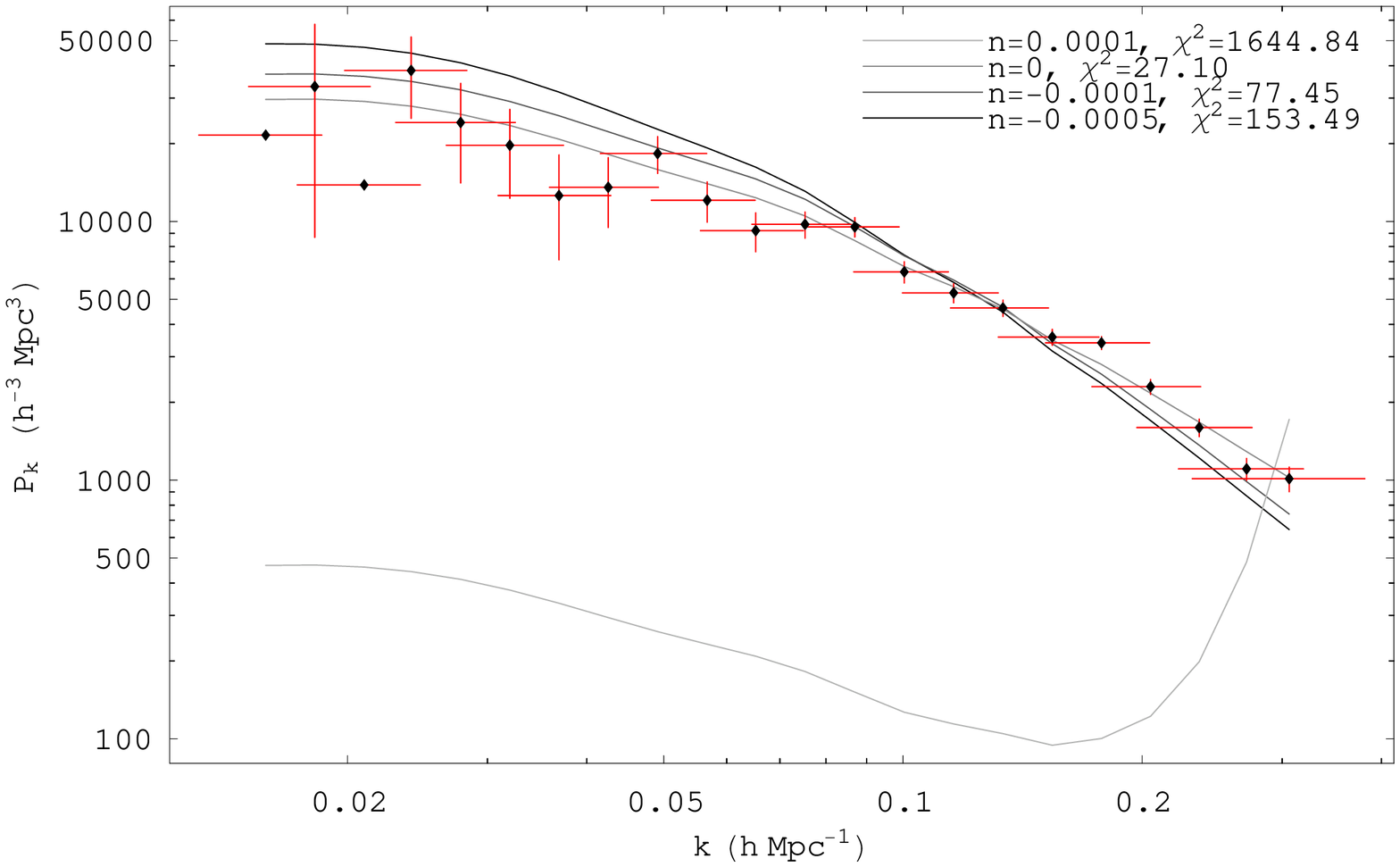}}
\caption{\label{card(n)powSpec}Power spectrum predicted by the Cardassian model versus SDSS data for several values of $n$. The curves are such that they best fit the data. The power spectrum predicted with the positive value $n=0.0001$ is totally excluded by the data.}
\end{figure}

We see clearly that the SDSS data do not allow for as large values of
$n$ as the supernovae data: at $2\,\sigma$, $n$ is limited to the
range $n\in\left[-1.5\times 10^{-5},8.9\times 10^{-6}\right]$. They
show that the Cardassian model must resemble a $\Lambda$CDM model very
closely to be in agreement with the large-scale structure observations
and the WMAP values for the cosmological parameters.
Figure~\ref{card(n)powSpec} compares the predictions of the Cardassian
model with the SDSS data and is in agreement with the results of
\cite{AmaElgMul04}. Note that negative values for $n$
  are not as strongly disfavoured as positive. But still only a very
  small deviation from zero in the negative direction is allowed.
\subsection{Constraining $n$ and the CDM density parameter $\Omega_{m0}$} \label{s42}
\begin{figure}[h]
\resizebox{\hsize}{!}{\includegraphics{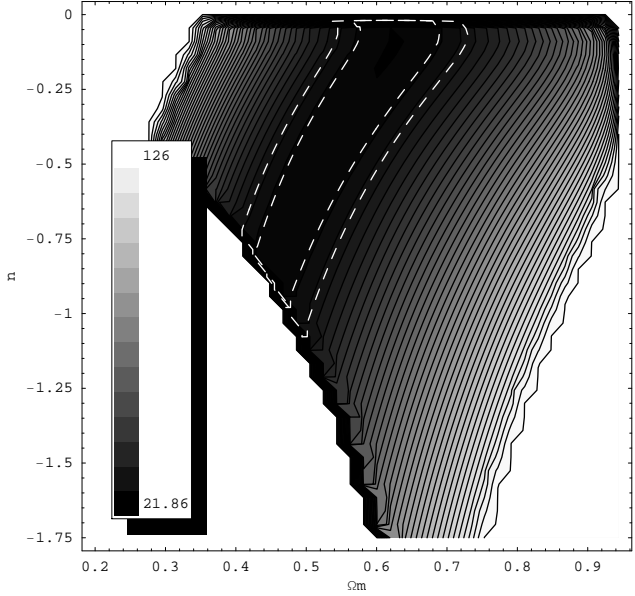}}
\caption{\label{card(nOm)coinLevel}$\chi^2$ and the $1$ and $2\,\sigma$ confidence contours (dashed lines) as functions of $n$ and $\Omega_{m0}$. We only present the values of $\chi^2$ smaller than $126$ to clarify the colour of the plot.}
\end{figure}

\begin{figure*}[h]
\centering
\includegraphics[width=17cm]{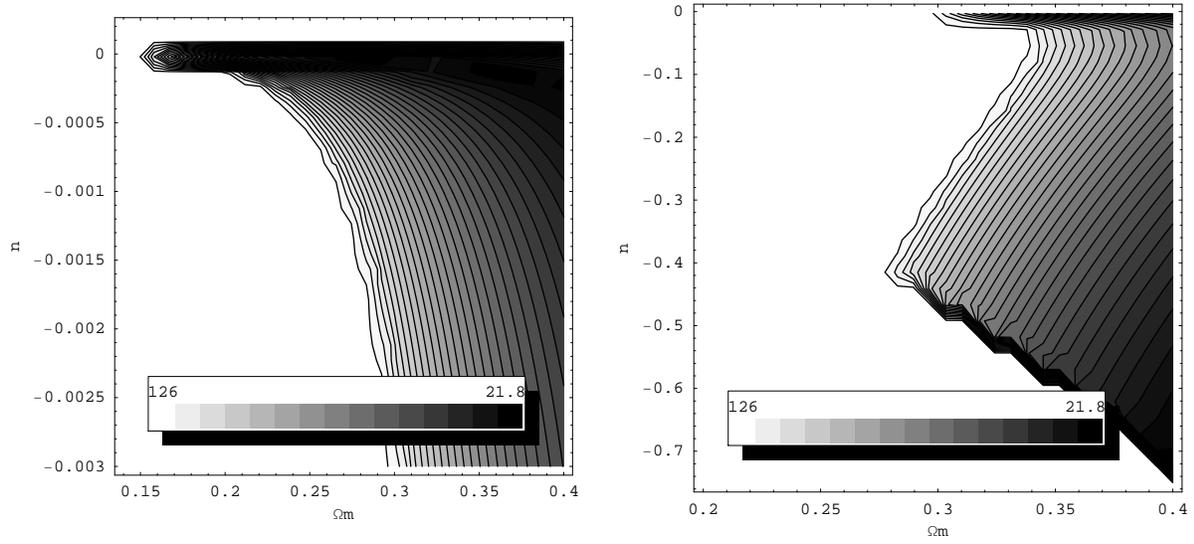}
\caption{\label{card(nOm)coinLevelFig123}Focus of figure \ref{card(nOm)coinLevel} in the region of $\Omega_{m0}$ allowed by WMAP data.}
\end{figure*}

\begin{figure}[h]
\centering
\includegraphics[width=4cm]{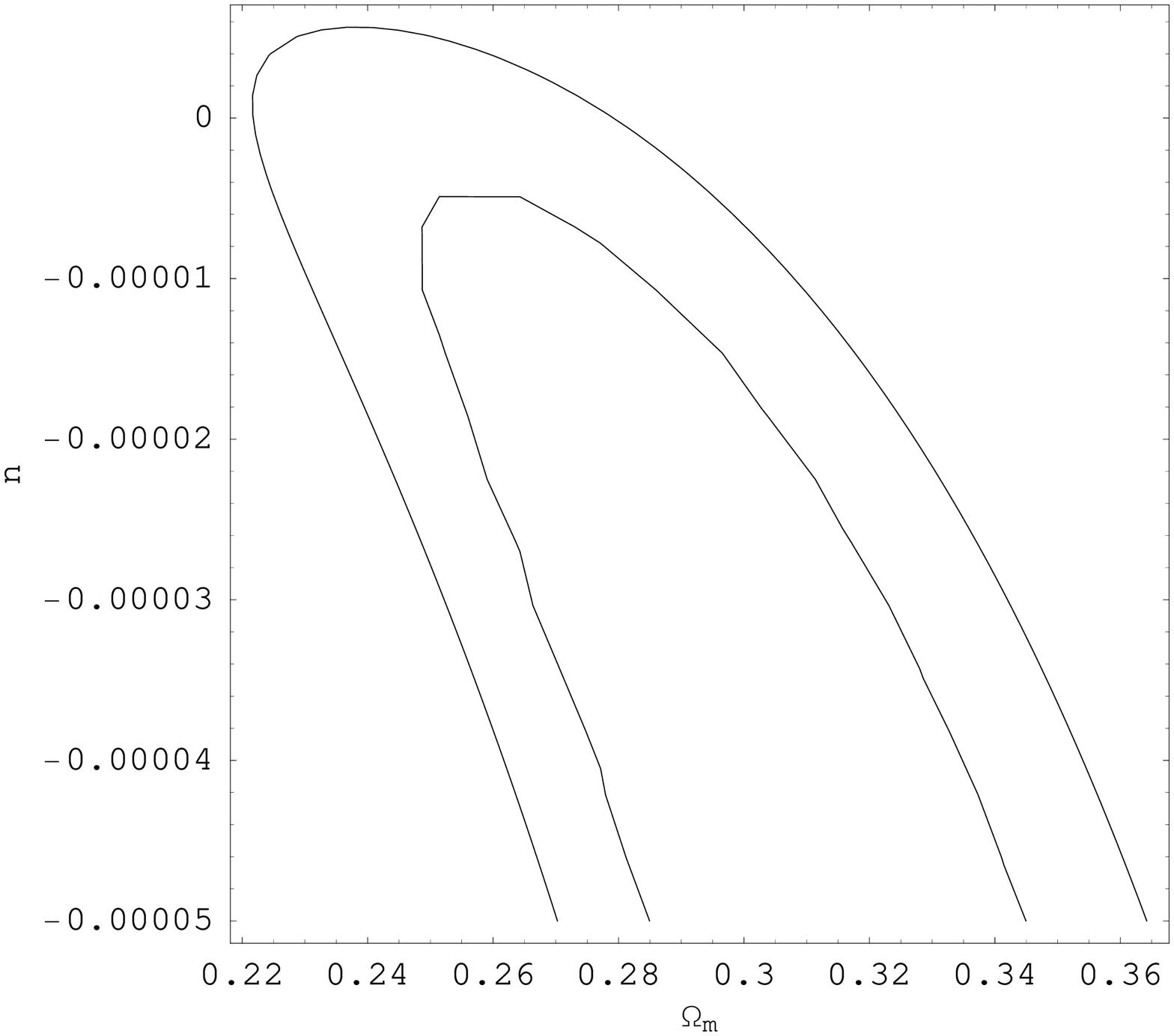}
\includegraphics[width=4cm]{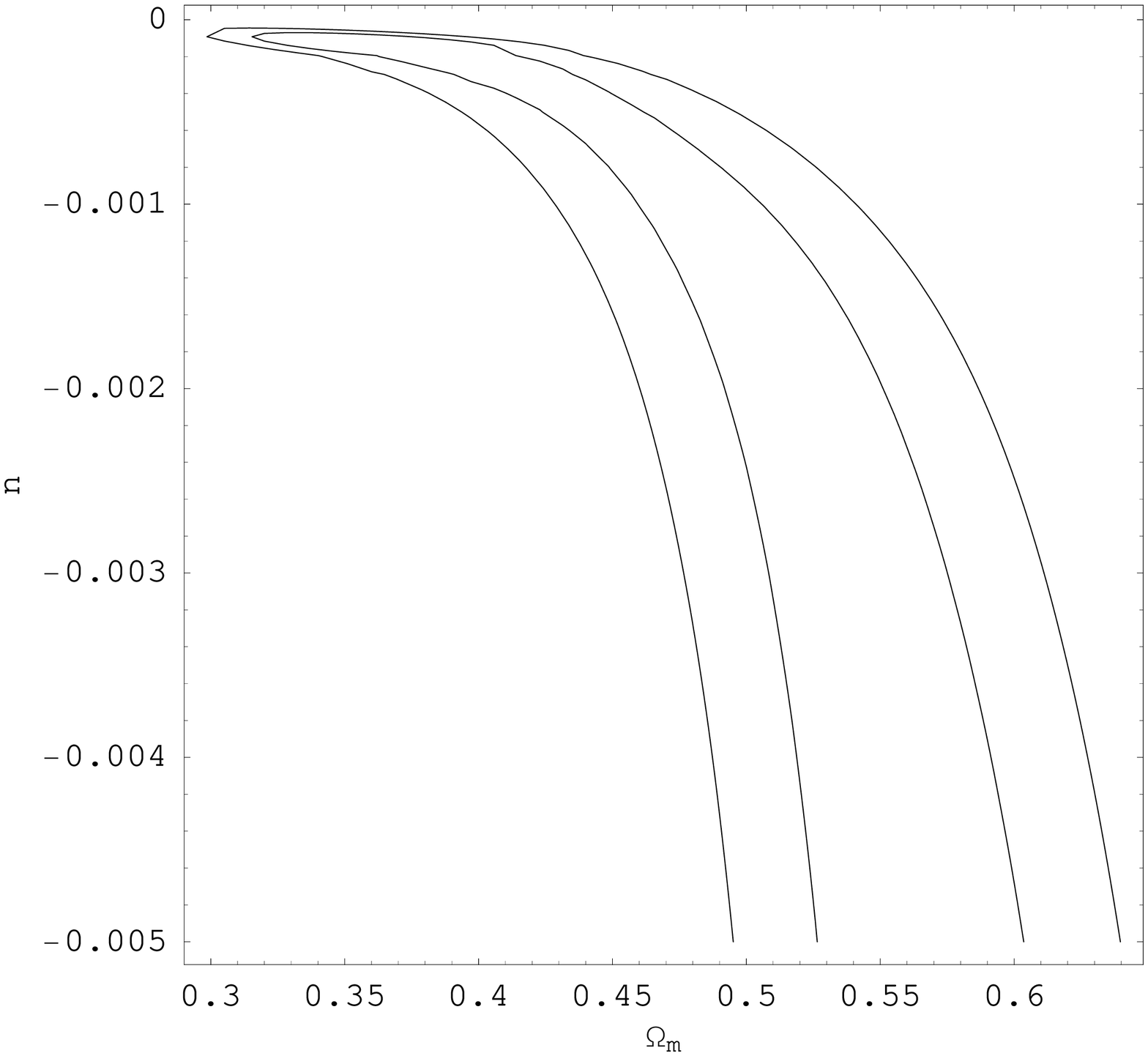}
\includegraphics[width=4cm]{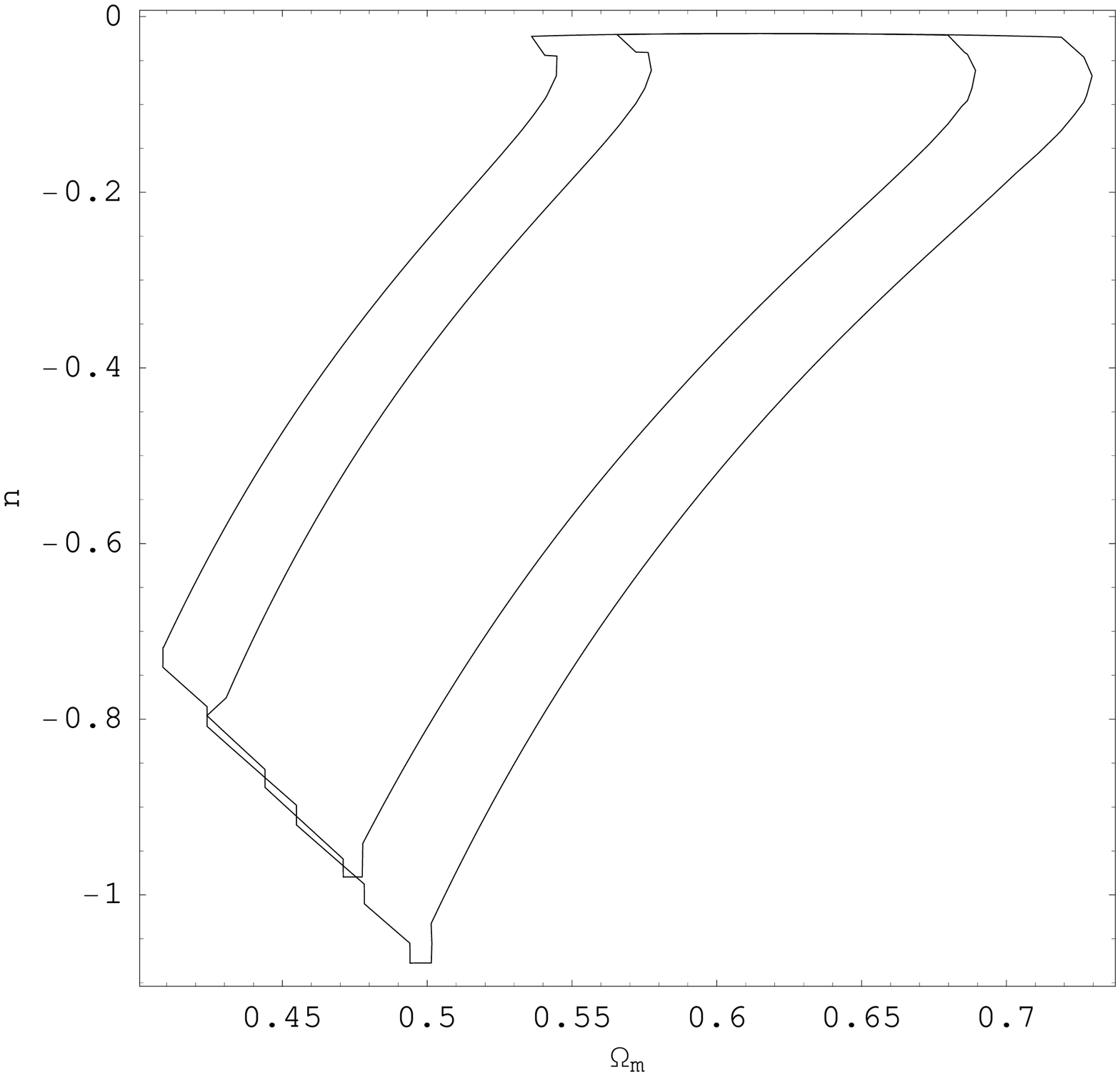}
\caption{\label{card(nOm)sigma}$1$ and $2\,\sigma$ confidence contours for several ranges of $n$. Only very small values of $n$ are possible when $\Omega_{m0}\approx0.224$, the value of the CDM density parameter in agreement with CMB data of WMAP.}
\end{figure}

This section and the following ones generalise the work of
\cite{AmaElgMul04}. We consider $n$ and $\Omega_{m0}$ to be free
parameters and keep the other cosmological parameters fixed to
$\Omega_{b0}=0.046$, $h=0.72$ and $n_s=1$. The minimal value for
$\chi^2$ is then $21.86$, which gives us $\chi^2_{DOF}=1.09$. This is
obtained when $\Omega_{m0}=0.62$ and $n=-0.10$. In
Fig.~\ref{card(nOm)coinLevel} we have plotted the  $\chi^2$ along with
the $1$ and $2\,\sigma$ confidence contours as functions of $n$ and
$\Omega_{m0}$. In Fig.~\ref{card(nOm)coinLevelFig123} one can find an
enlargement of this figure for the area $\Omega_{m0}<0.4$. The $1$ and
$2\,\sigma$ confidence levels for several ranges of values of $n$, in
particular for the small $n$, are plotted in the figure
\ref{card(nOm)sigma}. At $2\,\sigma$, the allowed ranges for the
parameters are $n\in\left[-1.1,6.3\times 10^{-6}\right]$ and
$\Omega_{m0}\in\left[0.22,0.73\right]$. Thus, we see
  that positive values for $n$ are strongly disfavoured. But the data
  do allow for negative values up to order of unity when treating only
  $n$ and $\Omega_{m0}$ as free parameters.

\begin{figure}[h]
\resizebox{\hsize}{!}{\includegraphics{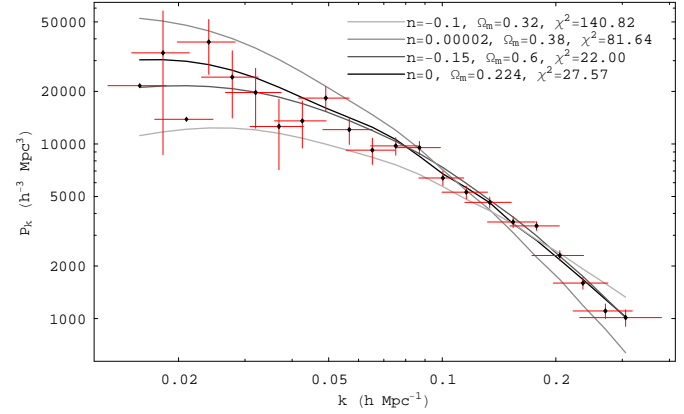}}
\caption{\label{card(nOm)powSpec}Some power spectra predicted by the Cardassian model for several values of $\Omega_{m0}$ versus the SDSS data.}
\end{figure}

In agreement with the previous section, for the fiducial value $\Omega_{m0}=0.224$, the Cardassian model fits the data only if $|n|<\mathcal{O}(10^{-5})$, that is: it is indistinguishable from a $\Lambda$CDM model. If we want the Cardassian model to be different from $\Lambda$CDM, one has to consider some unusual values of $\Omega_{m0}$ like $0.6$. Figure~\ref{card(nOm)powSpec} shows predictions of the Cardassian model for the power spectrum versus SDSS data for some values of $\Omega_{m0}$ and $n$.

\subsection{Constraining $n$ and the baryon density parameter $\Omega_{b0}$} \label{s43}
This time, we keep $\Omega_{c0}=0.73$, $h=0.72$ and $n_s=1$ fixed, but
treat $n$ and the baryon density parameter
  $\Omega_{b0}$ as free parameters when doing the fitting. The lowest $\chi^2$ value is now obtained for a negative value of $\Omega_{b0}$ (even with a prior on $\Omega_{b0}$) with $\chi^2=21.55$. This is clearly an unphysical value. However degeneracy also allows positive values of $\Omega_{b0}$ with reasonable $\chi^2$, e.g. $\chi^2=26.35$ when $\Omega_{b0}=0.038$ and $n=-1.1\times 10^{-5}$.

\begin{figure}[h]
\resizebox{\hsize}{!}{\includegraphics{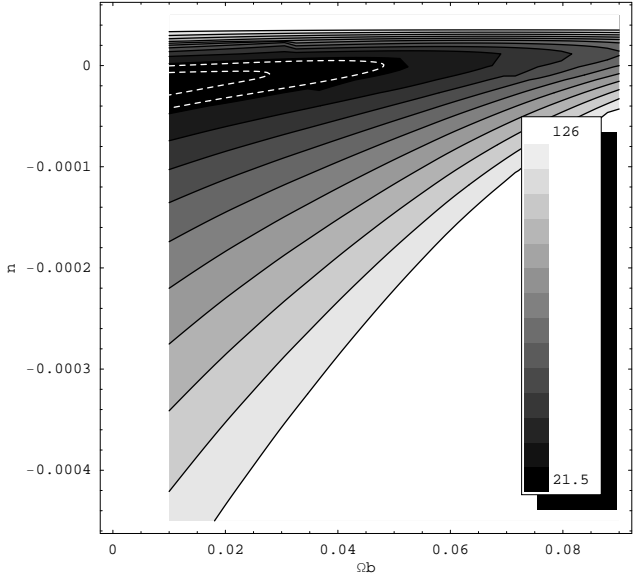}}
\caption{\label{card(nOb)coinLevelMulti}$\chi^2$ and the $1$ and $2\sigma$ confidence contours (dashed lines) as functions of $\Omega_{b0}$ and $n$. Only the small value of $n$ allow to fit the SDSS data.}
\end{figure}

\begin{figure}[h]
\resizebox{\hsize}{!}{\includegraphics{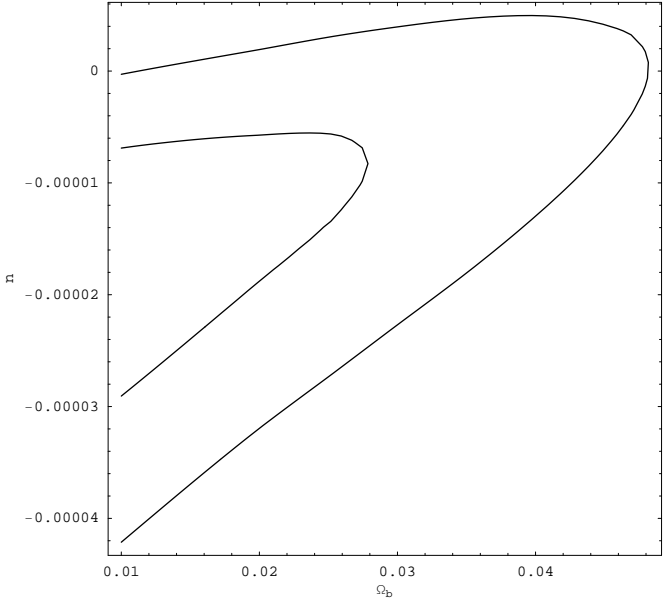}}
\caption{\label{card(nOb)sigma}The $1$ and $2\sigma$ confidence contours in the plane $(n,\Omega_{b0},)$.}
\end{figure}

Figure~\ref{card(nOb)coinLevelMulti} shows a plot of the $\chi^2$ and
the $1$ and $2\,\sigma$ confidence contours as functions of
$\Omega_{b0}$ and $n$. An enlargement of the confidence contours is
shown in figure~\ref{card(nOb)sigma} for the most relevant area of the
parameter space. The allowed ranges for the parameters at $2\,\sigma$
are $n\in\left[-4.3\times 10^{-5},5.3\times 10^{-6}\right]$ and
$\Omega_{b0}\in\left[0,0.048\right]$. Once again we find that the data
constrains $n$ very strongly. Just as when we constrained $n$ alone, we
find that the allowed range is smaller for positive than negative
values. But the constraint is still so strong that $n$ has to satisfy 
$|n|<10^{-5}$ in order for the Cardassian model to be in agreement
with the SDSS data.

\begin{figure}[h]
\resizebox{\hsize}{!}{\includegraphics{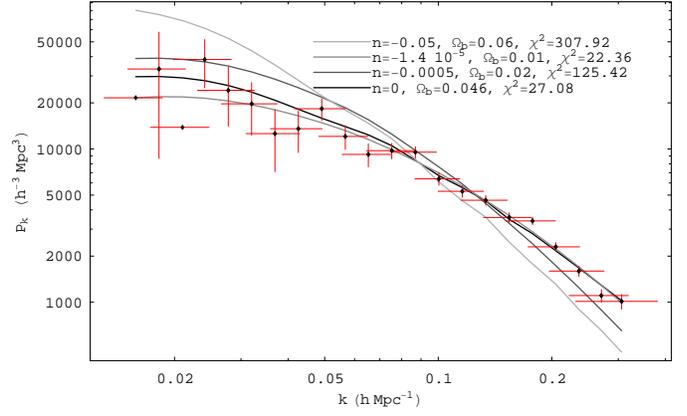}}
\caption{\label{card(nOb)powSpec}Some power spectra predicted by the
  Cardassian model for several values of $\Omega_{b0}$ versus the SDSS data.}
\end{figure}

Figure~\ref{card(nOb)powSpec} shows predictions of the Cardassian model for the power spectrum for some choices for $n$ and $\Omega_{b0}$ versus the SDSS data. We see from Fig.~\ref{card(nOb)sigma} that the data seem to prefer values for $\Omega_{b0}$ that are smaller than the WMAP value. But still, the allowed values of $n$ are so small that it leaves the Cardassian model virtually indistinguishable from a $\Lambda$CDM model.

\subsection{Constraining $n$ and the Hubble parameter $h$} \label{s44}
\begin{figure}[h]
\resizebox{\hsize}{!}{\includegraphics{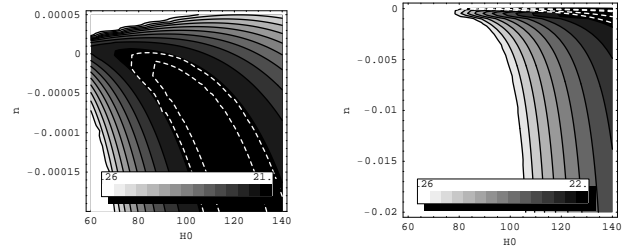}}
\caption{\label{card(nH0)coinLevelMulti}$\chi^2$ and the $1$ and $2$ sigma confidence contours (dotted line) as functions of $H_0$ and $n$.}
\end{figure}

\begin{figure*}[h]
\centering
\includegraphics[width=8.5cm]{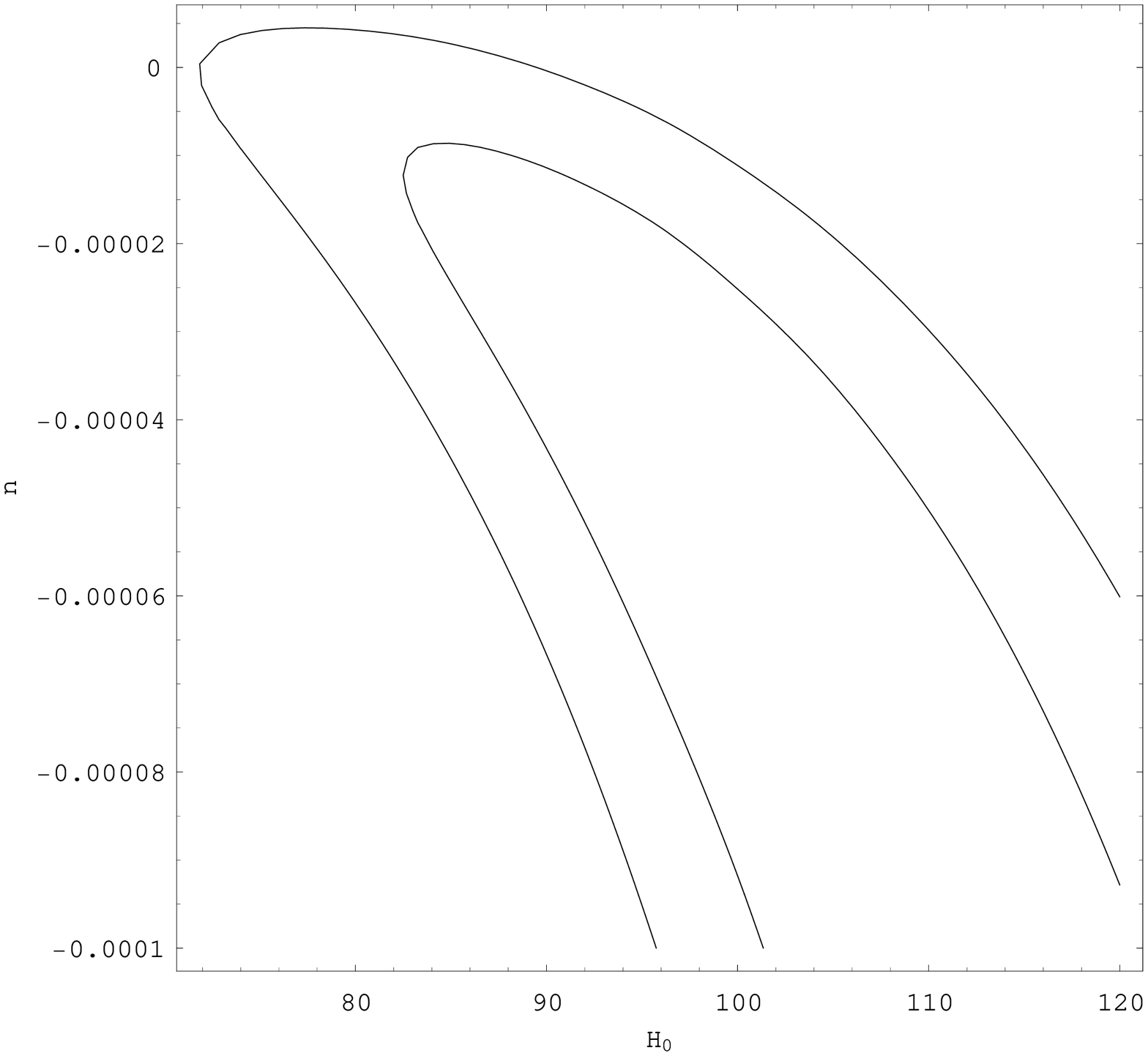}
\includegraphics[width=8.5cm]{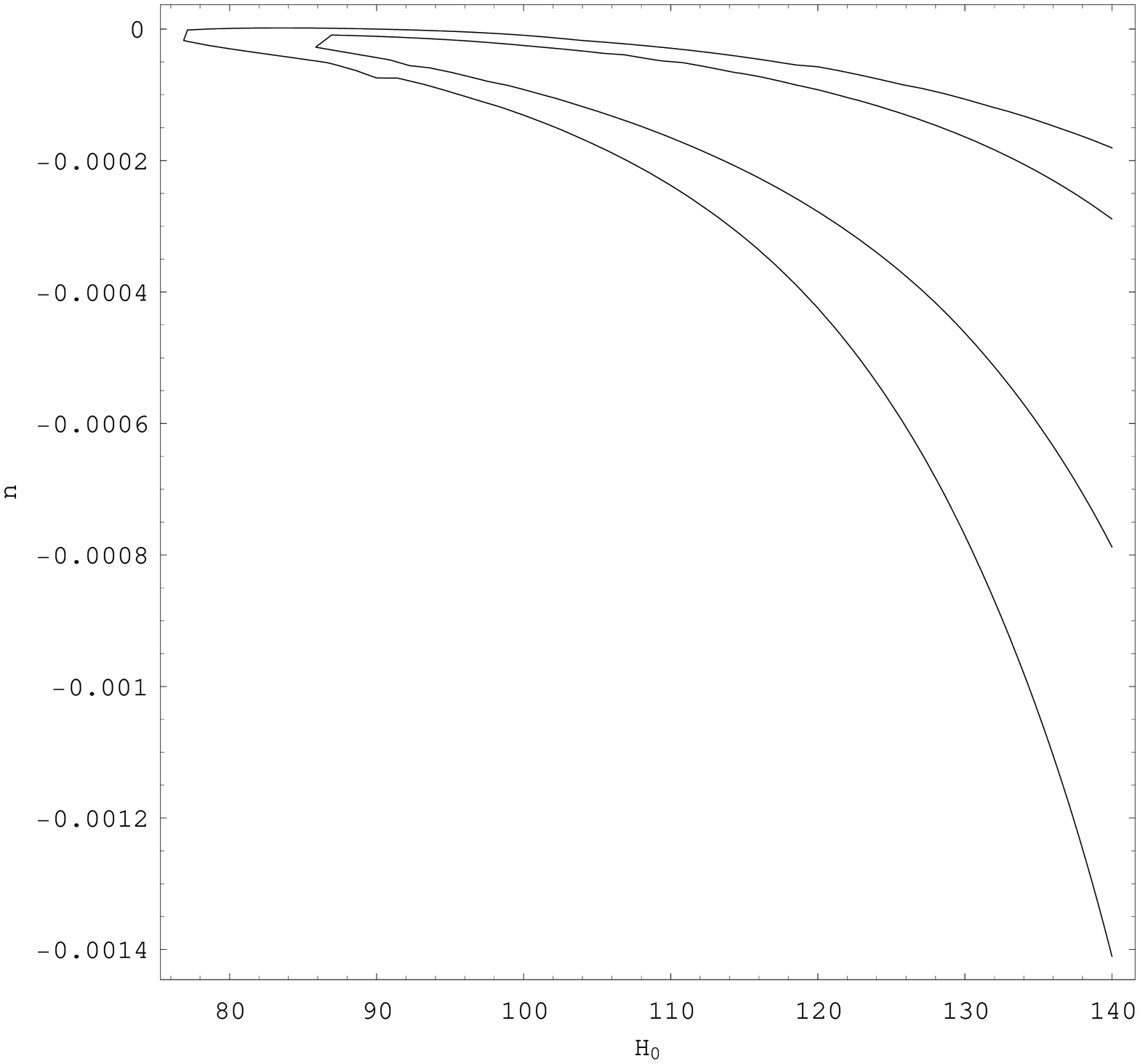}
\caption{\label{card(nH0)sigma}$\chi^2$ as a function of $H_0$ and $n$.}
\end{figure*}

Here, we treat $n$ and $h$ as free parameters in the
  fitting, and keep $\Omega_{m0}=0.224$, $\Omega_{b0}=0.046$ and
$n_s=1$ fixed. The best fit is now $\chi^2=21.02$ for $h=1.11$ and
$n=-9.8\times10^{-5}$ with $\chi^2_{DOF}=1.05$. Plots of the $\chi^2$
and the $1$ and $2\,\sigma$ confidence contours as functions of $h$
and $n$ can be found in figure~\ref{card(nH0)coinLevelMulti}. A couple
of enlargements of the confidence contours are plotted in
figure~\ref{card(nH0)sigma}. The parameters ranges allowed at
$2\,\sigma$ are $n<1.8\times 10^{-5}$ and $h>0.76$. We have not
calculated the lower and upper limits for respectively $n$ and $h$
since they are beyond $h=1.40$ (and then below $n=-8\times 10^{-4}$).
Thus, the data appear to place a tight limit on positive
  values for $n$, but not on negative values, at least if one doesn't
  put a prior on $h$.
Clearly, as the confidence contours show us, the SDSS data are not able to constrain the Hubble parameter very well. 

Note that the HST Key Project \citep{Freedman00} quotes a value for
the Hubble parameter of $h=0.72\pm0.8$, whereas the WMAP team quote a
value of $h=0.72\pm0.05$ when using WMAP data alone. If we demand $h$
to be around $0.7$ in order to be in agreement with these
measurements, we find that $|n|$ must be of order less than
$\mathcal{O}(10^{-5})$. Again, this implies that the Cardassian model
is indistinguishable from the $\Lambda$CDM model. The only way to make
it substantially different is to accept values for $h$ larger than
$1.0$, which is quite unrealistic with respect to the measurements
quoted above.  

\begin{figure}[h]
\resizebox{\hsize}{!}{\includegraphics{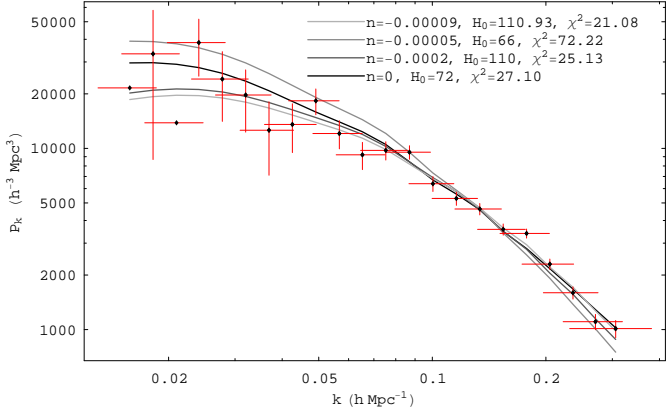}}
\caption{\label{card(nH0)powSpec}Some power spectra predicted by the Cardassian model for several values of $H_0$ versus the SDSS data.}
\end{figure}

Fig.~\ref{card(nH0)powSpec} show plots of the power spectra predicted by the Cardassian model for some choices for the parameters $n$ and $h$.
\subsection{Constraining $n$ and the spectral index $n_s$} \label{s45}
\begin{figure*}[h]
\centering
\includegraphics[width=17cm]{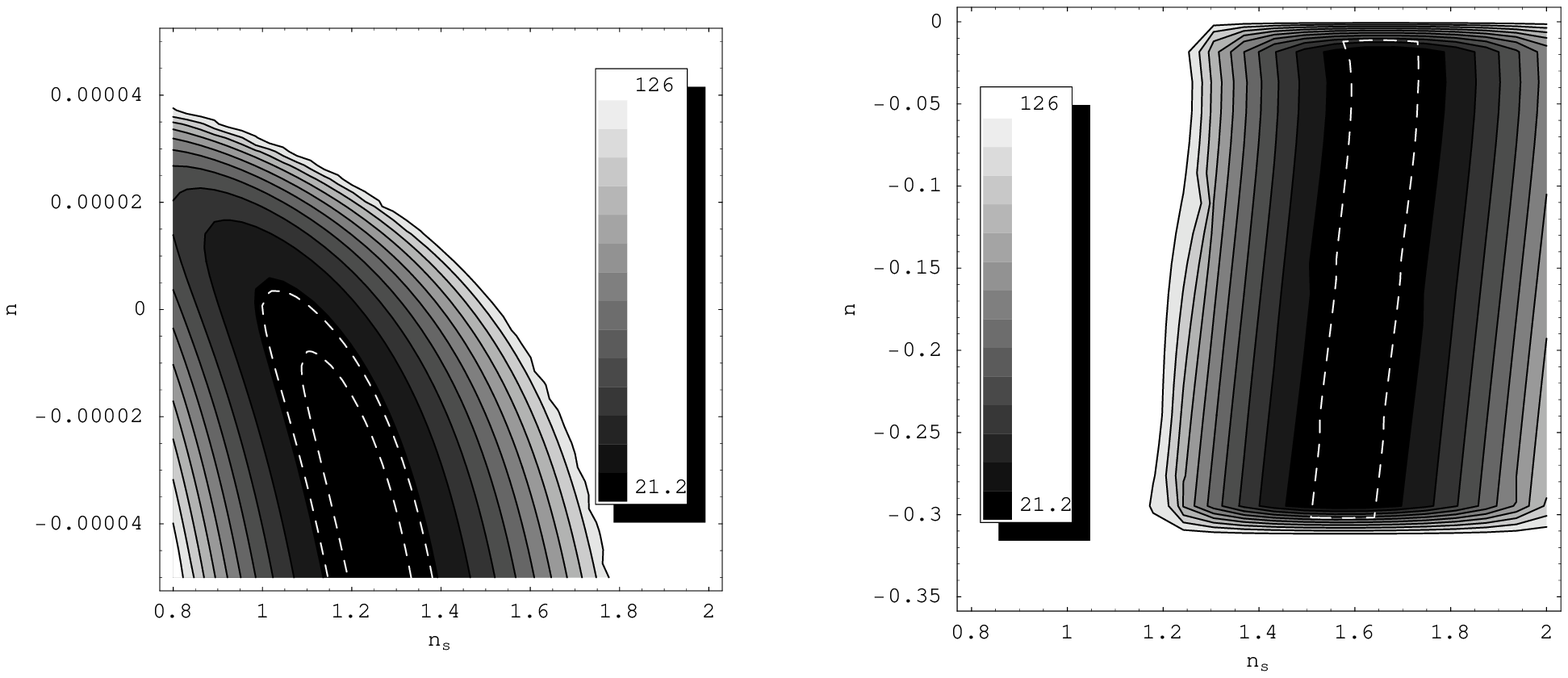}
\caption{\label{card(nk)coinLevel}$\chi^2$ and the $1$ and $2$ sigma confidence contours (dotted lines) as functions of $n_s$ and $n$. Note that in the second plot, the $1\sigma$ confidence contour is not visible at this scale.}
\end{figure*}

In all the previous sections we have chosen the Harrison-Zeldovich
power spectrum $P=k^{n_s}$ with $n_s=1$ as the initial power spectrum.
In this section we will consider $n_s$ as a free parameter along with
$n$. The other cosmological parameters are kept fixed with
$\Omega_{b0}=0.046$, $\Omega_{m0}=0.224$ and $h=0.72$. The best fit is
then $\chi^2=21.25$ with $\chi^2_{DOF}=1.06$ when $n_s=1.23$ and
$n=-4.0\times10^{-5}$. The $\chi^2$ and the $1$ and $2\,\sigma$
confidence contours are plotted in Fig.~\ref{card(nk)coinLevel} as
functions of $n_s$ and $n$. At $2\sigma$ the allowed ranges for the
parameters are $n\in\left[-0.3,4.2\times 10^{-6}\right]$ and
$n_s\in\left[1,1.73\right]$. Just as we saw in the constraints in
sections~\ref{s42} and \ref{s44}, positive values for $n$ are tightly
constrained, whereas relatively large negative values are allowed.

Thus, it is evident that the SDSS data do not tightly constrain $n_s$.
Indeed, we find that for $n_s\simeq1.6$, large values of $n<0$ are in
agreement with the SDSS data. Thus, if we accept such large values for
the spectral index, it is then possible to construct a Cardassian
model that differs considerably from $\Lambda$CDM and
  still agrees with the SDSS data. However, such values of $n_s$ are
not compatible with the value measured by WMAP. For the special value
$n_s=1$, i.e. a Harrison-Zeldovich spectrum, we find that $|n|$ has to be of order less than $\mathcal{O}(10^{-5})$ in order to be in agreement with the data. 

Once again, if we do not modify the usual Harrison-Zeldovich power spectrum with $n_s=1$, the Cardassian model is equivalent to a $\Lambda$CDM model. Fig.~\ref{card(nH0)powSpec} shows plots of the predicted power spectrum for the Cardassian model for several values of $n$ and $n_s$ along with the SDSS data.
\begin{figure}[h]
\resizebox{\hsize}{!}{\includegraphics{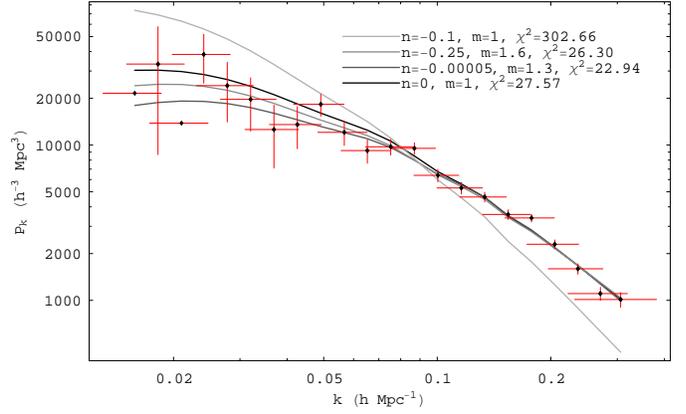}}
\caption{\label{card(nk)powSpec}Some power spectra predicted by the Cardassian model for several values of $n_s$ and $n$ versus the SDSS data.}
\end{figure}
\section{Conclusions} \label{s5}
The purpose of this work has been to constrain the parameters of
the power law Cardassian model in the fluid
interpretation by using large scale structure data from the SDSS
galaxy survey. We generalised the work of \cite{AmaElgMul04} by
treating additional cosmological parameters as free in addition to $n$
itself under a fitting to the
galaxy power spectrum. In our analysis, we looked at five different
fits to the SDSS galaxy power spectrum. In the first fit, repeating
the analysis of \cite{AmaElgMul04}, we fixed the four cosmological
parameters $\Omega_{m0}$ (CDM density), $\Omega_{b0}$ (baryon
density), $h$ (Hubble parameter) and $n_s$ (spectral index of the
initial spectrum), and treated only the Cardassian parameter $n$ as a
free parameter. One finds then that $|n|$ has to be of
order less that $10^{-5}$ in order to be in accordance with the data.
A similar conclusion was reached by \citet{Koivisto05}
  for the MPC model using the CMB power spectrum. These results are
  also similar to those of \citet{Sandvik02}, who showed that a
  generalised Chaplygin gas has to resemble the $\Lambda$CDM model
  closely to avoid oscillations or blow ups of the matter power spectrum.

In the remaining four fits, we allowed one additional parameter to
take values freely along with $n$, while keeping the
remaining ones fixed to their WMAP best fit value.
First, constraining $\Omega_{m0}$ simultaneously with
  $n$, we find that positive values for $|n|$ are tightly
constrained, but not the negative. While the $2\,\sigma$ upper limit
for n is of order $10^{-6}$, the lower limit allows for negative
values of order unity. But for such negative values the Universe
would have to contain an unsual amount of cold dark matter. If we
demand that the dark matter content be within the bounds given by
WMAP, we again find that $|n|$ has to be so small that the model
becomes indistinguishable from $\Lambda$CDM.

Next, varying $\Omega_{b0}$ and $n$ freely, we again
find that latter is strongly constrained both in the negative and
positive directions. More specifically, $|n|$ has to be of order
smaller than $10^{-5}$ to be in agreement with the data. Next,
varying $h$ and $n$ freely, we again find that
positive values for $n$ are tightly constrained, but negative values
are much less constrained. However, it turns out $h$ has to be larger
than $\sim1.4$ in order for $n$ to be smaller than
$-10^{-3}$. This is an unrealistically high value for the Hubble
parameter. Demanding a more realistic value for $h$ will again render
the Cardassian model indistinguishable from $\Lambda$CDM. 

Finally, allowing the spectral index $n_s$ and $n$ to
  vary freely, we find the same behaviour as in the last fit.
Positive values for $n$ are tightly constrained, while negative values
are not. But adding a reasonable prior on $n_s$ will restrict the
allowed negative values severely. In fact, $n_s$ has to be larger than
$\sim1.5$ in order for $n$ to be of order less than $-10^{-1}$ and still
agree with the SDSS data.
A summary of our results showing the best fit models can be found in table~\ref{tab1}.

Hence, unless one is ready to accept very unusual values for the
cosmological parameters considered here, the SDSS data force the
power law Cardassian model to be virtually
indistinguishable from an ordinary $\Lambda$CDM model.
It thus seems difficult to unify dark energy and dark
  matter via the power law Cardassian model.

Does this spell the end for the power law Cardassian
model -- at least in the fluid interpretation? A more general
treatment is probably needed to conclude this absolutely. For example,
one would ideally like to write a code like CMBFast for the Cardassian
model instead of using the $\Lambda$CDM model to generate the initial
perturbations. However, this work confirms that the SDSS data look
like a promising way to rule out this model. It shows that the galaxy
power spectrum impose much tighter constraints on the Cardassian model
than the supernova data.

Finally, we wish to stress that we have only
  considered the Cardassian model in the fluid interpretation. If
  an alternative interpretation were to be found, where perturbative
  calculation could be performed, a similar analysis might very well
  produce a different result.

\begin{table}[h]
\begin{center}
\begin{tabular}{llll}
\hline
Free parameters & Priors & $\chi^2_{DOF}$ & Best fit \\
\hline
$n$ & $n_s=1$, $\Omega_{m0}=0.224$, & $1.28$ & $n=-5.6\times 10^{-7}$ \\
	& $\Omega_{b0}=0.046$, $h=0.72$&	&	\\
$n$, $\Omega_{m0}$ & $n_s=1$, $\Omega_{b0}=0.046$, & $1.09$ & $\Omega_{m0}=0.62$, \\
	& $h=0.72$&	&	$n=-0.1$\\
$n$, $\Omega_{b0}$ & $n_s=1$, $\Omega_{c0}=0.73$,& $1.08$ & $\Omega_{b0}=-0.0022$,  \\
	& $h=0.72$&	&	$n=-2.0\times 10^{-5}$\\
$n$, $h$ & $n_s=1$, $\Omega_{m0}=0.224$,& $1.05$ & $h=1.11$, \\
	& $\Omega_{b0}=0.046$ &	&	$n=-9.0\times 10^{-5}$\\
$n$, $n_s$ & $\Omega_{m0}=0.224$, & $1.06$ & $n_s=1.23$,  \\
	& $\Omega_{b0}=0.046$, $h=0.72$&	&	$n=-4.0\times 10^{-5}$\\
\hline
\end{tabular}
\caption{\label{tab1}Best values of the cardassian parameters fitting the SDSS baryon spectrum}
\label{label}
\end{center}
\end{table}
\section*{Acknowledgments}
SF is supported by a Marie Curie Intra-European Fellowship Program of the Commission of the European Union (MEIF-CT-2005-515028) which is greatly appreciated. MA acknowledges support from the Norwegian Research Council through the project "Shedding Light on Dark Energy", grant 159637/V30. MA also wishes to thank \O{}ystein Elgar\o{}y for helpful comments.
\bibliographystyle{aa}

\begin{thebibliography}{26}
\expandafter\ifx\csname natexlab\endcsname\relax\def\natexlab#1{#1}\fi

\bibitem[{Amarzguioui et al.(2005)}]{AmaElgMul04}
Amarzguioui, M., et al. 2005, Combined constraints on Cardassian models from supernovae CMB and large-scale structure observations., JCAP, 0501:008

\bibitem[{Avelino et al.(2003)}]{Avelino02}
Avelino, P. P., et al. 2003, Alternatives to quintessence
model-building., Phys. Rev., D67:023511

\bibitem[{Beca et al.(2003)}]{Beca03}
Beca, L. M. G., et al. 2003, The Role of Baryons in Unified Dark
Matter Models. Phys. Rev., D67:101301

\bibitem[{Bento et al.(2005)}]{Bento05}
Bento, M. C., et al. 2005, Supernova constraints on models of dark
energy revisited. Phys. Rev., D71:063501

\bibitem[{Bento et al.(2006)}]{Bento06}
Bento, M. C., et al. 2006, Supernovae constraints on dark energy and
modified gavity models. J. Phys. Conf. Ser., 33:197 

\bibitem[{Chung \& Freese(2000)}]{Chung99}
Chung, D. J. H., \& Freese, K. 2000, Cosmological challenges in theories with extra dimensions and remarks on the horizon problem.
Phys. Rev., D61:023511

\bibitem[{Di Pietro \& Claeskens(2003)}]{PieCla03}
Di Pietro, E., \& Claeskens, J.-F. 2003, Future supernovae data and quintessence models. Mon. Not. Roy. Astron. Soc., 341:4

\bibitem[{Freedman et al.(2001)}]{Freedman00}
Freedman W. L., et al. 2001, Final Results from the Hubble Space Telescope Key Project to Measure the Hubble Constant. Astrophys. J., 553:47--72

\bibitem[{Freese \& Lewis(2002)}]{Freese02}
Freese, K., \& Lewis, M. 2002, Cardassian Expansion: a Model in which the Universe is Flat, Matter Dominated, and Accelerating. Phys. Lett., B540:1--8

\bibitem[{Frith(2004)}]{Frith03}
Frith, W. J. 2004, Constraints on Cardassian Expansion. Mon. Not. Roy. Astron. Soc., 348:916

\bibitem[{Gondolo \& Freese(2003)}]{Gondolo02}
Gondolo, P., \& Freese, K. 2003, Fluid interpretation of Cardassian expansion. Phys. Rev., D68:063509

\bibitem[{Gong \& Duan(2004)}]{Gong04}
Gong, Y.-G., \& Duan, C.-K. 2004, Supernova constraints on alternative models to dark energy. Mon. Not. Roy. Astron. Soc., 352:847

\bibitem[{Knop et al.(2003)}]{Knop03}
Knop, R. A., et al. 2003, New Constraints on $\Omega_M$, $\Omega_\Lambda$, and w from an Independent Set of Eleven High-Redshift Supernovae Observed with HST. Astrophys. J., 598:102

\bibitem[{Koivisto et al.(2005)}]{Koivisto05}
Koivisto, T., et al. 2005, The CMB spectrum in Cardassian models.,
Phys. Rev., D71:064027.

\bibitem[{Lazkoz et al.(2005)}]{Lazkoz05}
Lazkoz, R., Nesseris, S., \& Perivolaropoulos, L. 2005, Exploring
Cosmological Expansion Parametrizations with the Gold SnIa Dataset.,
JCAP., 0511:010

\bibitem[{Lyth \& Stewart(1990)}]{LytSte90}
Lyth, D. H., \& Stewart, E. D. 1990, The evolution of density perturbations in the universe., Astrophys. J., 361:343--353

\bibitem[{Nesseris \& Perivolaropoulos(2004)}]{Nesseris04}
Nesseris, S., \& Perivolaropoulos, L. 2004, A comparison of cosmological models using recent supernova data. Phys. Rev., D70:043531

\bibitem[{Padmanabhan(1993)}]{Pad93}
Padmanabhan, T. 1993, Structure Formation in the Universe., Cambridge University Press, Cambridge

\bibitem[{Perlmutter et al.(1999)}]{Per99}
Perlmutter, S., et al. 1999, Measurements of ${\Omega}$ and ${\Lambda}$ from 42 {H}ight-{R}edshift {S}upernovae.
 Astrophys. J., 517:565--586

\bibitem[{Riess et al.(1998)}]{Riess98}
Riess, A. G. et al. 1998, Observational evidence from supernovae for an accelerating universe
 and a cosmological constant. Astron. J., 116:1009--1038

\bibitem[{Riess et al.(2004)}]{Rie04}
Riess, A. G., et al. 2004, Type Ia Supernova Discoveries at $z>1$ From the Hubble Space Telescope: Evidence for Past Deceleration and Constraints on Dark Energy Evolution. Astrophys. J., 607:665--687

\bibitem[{Sandvik et al.(2004)}]{Sandvik02}
Sandvik, H., et al. 2004, The end of unified dark matter?. Phys. Rev., D69:123524

\bibitem[{Seljak \& Zaldarriaga(1996)}]{SelZal96}
Seljak, U. \& Zaldarriaga, M. 1996, A Line of Sight Approach to Cosmic Microwave Background Anisotropies., Astrophys.J., 469:437--444

\bibitem[{Sen \& Sen(2003)}]{Sen02}
Sen, S., \& Sen, A. A. 2003, Observational Constraints on Cardassian Expansion. Astrophys. J., 588:1--6

\bibitem[{Spergel et al.(2003)}]{Spergel03}
Spergel, D. N., et al. 2003, First Year Wilkinson Microwave Anisotropy Probe (WMAP) Observations: Determination of Cosmological Parameters.
 Astrophys. J. Suppl., 148:175

 \bibitem[{Spergel et al.(2006)}]{Spergel06}
Spergel, D. N., et al. 2006, Wilkinson Microwave Anisotropy Probe (WMAP) three year results: Implications for cosmology.
 Submitted to ApJ.

\bibitem[{Szydlowski et al.(2005)}]{Szyd04}
Szydlowski, M., \& Czaja, W. 2005, Modified Friedmann Cosmologies -
theory \& observations. Annals Phys., 320:251

\bibitem[{Szydlowski et al.(2006)}]{Szyd05}
Szydlowski, M. \& Godlowski, W. 2006, Which cosmological models - with
dark energy or modified FRW dynamics?, Phys. Lett., B633:427

\bibitem[{Tegmark et al.(2004)}]{Tegmark03a}
Tegmark, M., et al. 2004, Cosmological parameters from SDSS and WMAP. Phys. Rev., D69:103501

\bibitem[{Tegmark et al.(2004)}]{Teg03}
Tegmark M., et al. 2004, The 3D power spectrum of galaxies from the SDSS. Astrophys. J., 606:702--740

\bibitem[{Tonry et al.(2003)}]{Tonry03}
Tonry, J. L., et al. 2003, Cosmological Results from High-z Supernovae. Astrophys. J., 594:1--24

\bibitem[{Wang et al.(2003)}]{Wang03}
Wang, Y., et al. 2003, Future type IA supernova data as tests of dark energy from modified Friedmann equations. Astrophys. J., 594:25--32

\bibitem[{Zhu et al.(2004)}]{Zhu04}
Zhu, Z.-H. , et al. 2004, Observational constraints on cosmology from modified Friedmann equation. Astrophys. J., 603:365--370
\end{thebibliography}

\end{document}